\documentclass[12pt, oneside]{article}   	
\usepackage[margin=1in]{geometry}
\usepackage{graphicx}							
\usepackage{amssymb}
\usepackage{amsmath}
\usepackage{amsthm}
\usepackage{tikz}
\usetikzlibrary{patterns, shapes.geometric, positioning, arrows}
\usepackage{booktabs}
\usepackage{breqn}
\usepackage{setspace}
\usepackage{algorithm,algorithmic}
\usepackage{authblk}
\usepackage{setspace}

\tikzstyle{block} = [rectangle, draw, fill=blue!20, 
    text width=20em, text centered, rounded corners, minimum height=4em]
\tikzstyle{line} = [draw, -latex']

\newtheorem{assumption}{Assumption}
\newtheorem{theorem}{Theorem}[section]

\newtheorem{lemma}[theorem]{Lemma}

\usepackage{url}
\usepackage{breakurl}
\usepackage[breaklinks]{hyperref}
\usepackage{apacite}
\let\cite\shortcite
\let\citeA\shortciteA

\title{Brief Article}
\author{Lamar Hunt III, Irene B. Murimi, Jodi B. Segal, Marissa J. Seamans, \\ Daniel O. Scharfstein, Ravi Varadhan}

\begin{document}
\title{Brand vs. Generic: Addressing Non-Adherence, Secular Trends, and Non-Overlap}
\maketitle

\section*{Abstract}
While generic drugs offer a cost-effective alternative to brand name drugs, regulators need a method to assess therapeutic equivalence in a post market setting. We develop such a method in the context of assessing the therapeutic equivalence of immediate release (IM) venlafaxine, based on a large insurance claims dataset provided by OptumLabs\textsuperscript{\textregistered}. To properly address this question, our methodology must deal with issues of non-adherence, secular trends in health outcomes, and lack of treatment overlap due to sharp uptake of the generic once it becomes available. We define, identify (under assumptions) and estimate (using G-computation) a causal effect for a time-to-event outcome by extending regression discontinuity to survival curves. We do not find evidence for a lack of therapeutic equivalence of brand and generic IM venlafaxine.

Key Words: G-computation, Generic Drugs,  Regression Discontinuity, Temporal Confounding, Therapeutic Equivalence, Time-Varying Confounding, Venlafaxine.


\bibliographystyle{apacite}

\section{Introduction}

Are brand and generic drugs equivalent? And what does equivalent mean? \citeA{strom1987generic} discusses three types of equivalence: chemical, biological, and therapeutic. Two drugs are chemically equivalent if they have the same quantity of active ingredient, even if other inactive ingredients differ. They are biologically equivalent (or \textit{bioequivalent}) if they are chemically equivalent and also have a comparable degree of bioavailability when administered to the same patient, where bioavailability is a measure of how well a drug can deliver its active ingredient to the site of action. Finally, they are therapeutically equivalent if they are bioequivalent and also have the same clinical outcomes when administered to the same patient \cite{strom1987generic}.

Since the passage of the Hatch-Waxman Act in 1984, generic drug producers do not need to perform randomized trials to market a generic drug as therapeutically equivalent to brand \cite{mossinghoff1999overview}. All that is required is a proof of bioequivalence. This Act significantly reduced the cost of producing generic drugs. Subsequently, the percentage of all dispensed drugs in the US which are generics grew from $18.6\%$ in $1984$ to $74.5\%$ in $2009$, and to $86.2\%$  in 2017 \cite{berndt2011brand, drugTrend}. This has undoubtedly benefited patients in the form of lower drug costs for conditions that are treatable with a generic drug; but the result is that many patients are now taking drug formulations whose clinical outcomes have never been directly tested in randomized clinical trials. 

As noted in \citeA{strom1987generic}, bioequivalence does not \textit{a priori} entail therapeutic equivalence, because the inactive ingredients that differ between the two drugs may interact with how the drug performs even after it has bound to the site of action. Furthermore, \citeA{bate2016generics} discuss several concerns about how the FDA evaluates whether two drugs are bioequivalent. For example, the FDA does not regularly perform independent tests of bioequivalence, and there have been several cases of companies fraudulently altering data to establish bioequivalence. A notable example is the generics producer Ranbaxy, which had to pay a \$500 million settlement and remove over 30 products from the market after widespread fraud was uncovered by a whistleblower within the company \cite{kay2013indian}. 

Another concern addressed by \citeA{bate2016generics} is that the comparisons of bioavailability often required by the FDA do not address the rate of absorption of the drug. For example, although the antidepressant Budeprion XL 150 mg, marketed by generics producer Teva, was shown to be bioequivalent to its brand version, Welbutrin XL 150 mg, the mean plasma concentrations across time of the generic have a different profile from brand. Despite the fact that this could lead to different therapeutic effects, the FDA considers Budeprion XL 150 mg to be bioequivalent to brand. What's more, the FDA allowed these data to be extrapolated as evidence in support of the bioequivalence of Budeprion XL 300 mg to Wellbutrin 300 mg. The FDA eventually rescinded this decision when, in response to concern about adverse effects experienced by patients switching to generic, they performed their own independent analysis showing ``Budeprion XL 300 mg tablets fail to release bupropion into the blood at the same \textit{rate} and to the same extent as Wellbutrin XL 300 mg" (emphasis added) \cite{fdaUpdate}. This inconsistency is a concern for prescribers trying to infer therapeutic equivalence from claims of bioequivalence alone.

In the absence of clinical trials, regulators need a method to accurately monitor for therapeutic differences in a post market setting. The goal of this paper is to develop such a method, and apply it to the antidepressant venlafaxine as a test case. To help define a relevant target parameter, it is useful to consider an ideal randomized clinical trial for assessing therapeutic equivalence (see \citeA{hernan2006estimating} for a discussion of the relationship between ideal trials and causal inference from observational data).

\subsection*{The Ideal Trial}
Suppose we want to test whether generic venlafaxine differs from its brand name counterpart, Effexor, with respect to some therapeutically relevant outcome in patients being treated for a major depressive episode. Ideally we would enroll patients who are currently experiencing a major depressive episode but have not been treated recently with any antidepressants, and randomize them to receive either continuous treatment with brand or generic venlafaxine. Although patients cannot be forced to take their medication, we can ensure that the drug is continuously available by providing them with a 30 day supply of either brand or generic every 30 days. Nothing else about the two treatment arms would differ, including the frequency with which they are provided with drugs, the days of drug supplied, and the financial incentives (or disincentives) associated with receiving the drugs. In other words, the two idealized interventions would be (1) patients are provided with a 30 day supply of brand every 30 days at a fixed out of pocket cost $p$, and (2) patients are provided with a 30 day supply of generic every 30 days at the same fixed out of pocket cost $p$. Typically it takes 3 months for patients to respond to treatment, plus another 6 months of continuation for them to stabilize, so we would choose a follow-up period of 9 months during which to observe the therapeutic outcomes of interest \cite{shelton2001steps, pringsheim2016stopping}. 


Our goal is to estimate the effect in this ideal randomized trial using observational data. Specifically, we develop a method to compare brand vs. generic venlafaxine under these two interventions using insurance claims data, where the outcome of interest is failure time. We define a failure event to be a composite of the following: (1) {\em treatment change}, defined by treatment with an antidepressant other than venlafaxine, an anti-psychotic, lithium, or electroconvulsive therapy (ECT), whether or not venlafaxine is continued; (2) {\em clinical progression}, defined by a suicide related clinical encounter, hospitalization or emergency department visit; and (3) death. Our target parameter is the difference in survival curves under each intervention.

Identification of the target parameter from observational data requires addressing non-adherence, secular trends, and non-overlap in treatment. Non-adherence can result in bias due to \textit{time-varying confounding} and \textit{informative censoring}, while secular trends can result in bias due to \textit{temporal confounding}. Time-varying confounding occurs when a time-varying exposure status (e.g., whether a patient is filling a prescription, and whether it is for brand or generic) impacts, and is impacted by, a time-varying covariate that also impacts the outcome \cite{robins2000marginal}. Informative censoring occurs when premature termination of enrollment is associated with the outcome \cite{campigotto2014impact}. Temporal confounding occurs when treatment initiation time impacts both the outcome (due to a secular trends in the outcome) and treatment status \cite[p.~1414]{rom2007environmental}. Finally, non-overlap in treatment with brand and generic results in a \textit{positivity} violation, making it impossible to adjust for temporal confounding using methods like standardization or inverse propensity score weighting \cite{westreich2010invited}.

While non-adherence is a potential source of bias in any study comparing efficacy of brand and generic drugs (and therefore therapeutic equivalence), secular trends and treatment non-overlap are problems in a broad variety of studies comparing brand to generic drugs. Previous studies comparing brand to generic drugs have failed to adjust for these. For example, \citeA{jackevicius2016comparative} only adjust for baseline variables using propensity scores, despite temporal confounding being a potential source of bias in their data. In this paper we discuss how to address all of these issues and apply the methodology to venlafaxine. Our solution involves extending the framework of Regression Discontinuity \cite{imbens2008regression} to entire survival curves that are themselves estimated using G-computation \cite{robins1986new}.

\section{Data}

This study involves a retrospective analysis of claims data from the OptumLabs\textsuperscript{\textregistered} Data Warehouse (OLDW), which includes de-identified claims data for privately insured and Medicare Advantage enrollees in a large, private, U.S. health plan. The database contains longitudinal health information on enrollees, representing a diverse mixture of ages, ethnicities and geographical regions across the United States. The health plan provides comprehensive full insurance coverage for physician, hospital, and prescription drug services \cite{optum}.

The exposure information is inferred from insurance claims for prescription fills of both brand and generic forms of 3 different formulations of venlafaxine: immediate release tablets (IM), extended release capsules (ERC), and extended release tablets (ERT). We observe the calendar time of the fill, the formulation that was filled, whether it was brand or generic, the days supply of the fill, and the out of pocket cost associated with the fill. Note that we are treating the out of pocket cost associated with filling venlafaxine as part of the joint intervention of interest, since the cost associated with each prescription fill is used to define the interventions in the ideal trial above.

The outcome information is inferred from insurance claims for other medical procedures and events that would constitute a failure event, such as a psychiatric hospitalization, a prescription fill of an anti-psychotic, or a suicide attempt. Suicide related events are inferred from the diagnosis code on the claim. 

We exclude patients who have received treatment with antidepressants, anti-psychotics, lithium, valproate, lamotrigine or ECT any time during the 180 days prior to their first prescription fill of venlafaxine. We only included those who initiated on the immediate release formulation. The eligible sample included $42$,$847$ patients. We excluded $381$ patients who initiated on brand after generic became available, and $87$ patients missing any baseline data, for a total sample size of $42$,$379$.

Baseline covariates include age at initiation on venlafaxine (\texttt{age}), \texttt{sex}, \texttt{race}, socioeconomic status (\texttt{ses}), Charlson co-morbidity index (\texttt{cci}), and number of outpatient visits in the past 180 days at baseline (\texttt{out}). Socioeconomic status was defined using the decile of the median household income for the patient's zip-code, calculated using census data from the years 2000, 2011, and 2014--whichever was closest to the date of initiation. Time varying covariates were computed daily, and include the number of prescriptions on hand excluding venlafaxine on day $t$, $\texttt{rxb}_{\texttt{t}}$; and a running total of out of pocket pharmacy expenditures, excluding those for venlafaxine, from the past 180 days measured on day $t$, $\texttt{oop}_{\texttt{t}}$. 

\begin{figure}[!h]
  \includegraphics[width=\linewidth]{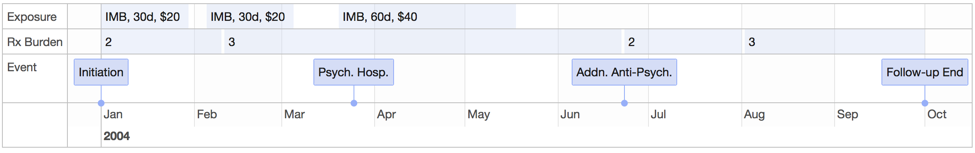}
  \caption{Hypothetical follow-up of a patient initiating on IM Brand.}
  \label{fig:PatientTimeline}
\end{figure}

In Table \ref{table:tableOne} we present a comparison of baseline values for brand and generic users. In Figure \ref{fig:PatientTimeline} we present a hypothetical follow-up of a patient initiating on IM Brand (IMB) on January 1st, 2004. The patient filled two 30 day prescriptions for IM Brand in January and February both at a cost of \$20, and one in March for 60 days at a cost of \$40. The non-shaded exposure days indicate the patient has no prescription for any form of venlafaxine on hand. The psychiatric hospitalization in March is a failure event, as is the addition of an anti-psychotic in late June. In this case, the patient's failure time would correspond to the date of psychiatric hospitalization, because it occurs first. October 1st, 9 months after initiation, marks the end of this patient's follow-up. We also show hypothetical values of \texttt{rxb} on days that it changes.

\begin{table}[!h]
\centering 
$\begin{array}{ | l | c | c | }
\hline
	& \text{IM Brand} & \text{IM Generic }\  \\ \hline
	\text{n} & 16619 & 25760 \\ \hline
	\texttt{rxb} (\%) & \  & \  \\
	\multicolumn{1}{|r|}{0} & 27.6 & 26.1 \\ 
	\multicolumn{1}{|r|}{1} & 26.0 & 25.7 \\ 
	\multicolumn{1}{|r|}{2} & 18.3 & 17.8 \\ 
	\multicolumn{1}{|r|}{3} & 11.1 & 11.3 \\ 
	\multicolumn{1}{|r|}{>3} & 17.1 & 19.1 \\ \hline
	\texttt{race} (\%) & \  & \  \\ 
	\multicolumn{1}{|r|}{\text{Asian}} & 0.8 & 1.1 \\ 
	\multicolumn{1}{|r|}{\text{Black}} & 2.7 & 4.6 \\ 
	\multicolumn{1}{|r|}{\text{Hispanic}} & 4.7 & 7.4 \\ 
	\multicolumn{1}{|r|}{\text{Multiple}} & 4.9 & 11.3 \\ 
	\multicolumn{1}{|r|}{\text{Unknown}} & 27.1 & 4.6 \\ 
	\multicolumn{1}{|r|}{\text{White}} & 59.8 & 71.0 \\ \hline
	\texttt{sex} = \text{male} (\%) & 24.8 & 23.1 \\ \hline
	\texttt{age} (\text{mean (sd)}) & 43.8 (11.2) & 44.7 (11.6) \\ \hline
	\texttt{cci} = 0 (\%) & 75.8 & 76.9 \\ \hline
	\texttt{out} (\%) & \  & \  \\
	\multicolumn{1}{|r|}{0} & 14.0 & 9.2 \\ 
	\multicolumn{1}{|r|}{1} & 80.9 & 85.1 \\ 
	\multicolumn{1}{|r|}{>1} & 5.1 & 5.7 \\ \hline
	\texttt{ses} (\%) & \  & \  \\ 
	\multicolumn{1}{|r|}{0} & 2.0 & 2.7 \\ 
	\multicolumn{1}{|r|}{1} & 4.0 & 6.6 \\ 
	\multicolumn{1}{|r|}{2} & 5.3 & 8.2 \\ 
	\multicolumn{1}{|r|}{3} & 6.1 & 8.8 \\ 
	\multicolumn{1}{|r|}{4} & 7.7 & 9.0 \\ 
	\multicolumn{1}{|r|}{5} & 8.8 & 9.8 \\ 
	\multicolumn{1}{|r|}{6} & 10.0 & 10.7 \\ 
	\multicolumn{1}{|r|}{7} & 16.8 & 13.0 \\ 
	\multicolumn{1}{|r|}{8} & 20.3 & 15.6 \\ 
	\multicolumn{1}{|r|}{9} & 18.8 & 15.8 \\ \hline
	\texttt{oop} (\text{mean (sd)}) & 137.7 (180.14) & 167.1 (252.8) \\ \hline
\end{array}$
\caption{Baseline Covariate Summaries} 
\label{table:tableOne} 
\end{table}

\section{Notation}
Let $T$ denote failure time (observed if the patient is not censored), with $S_t = I\{T \leq t\}$; $C$ denote censoring time, with  $C_t = I\{C \leq t\}$; $Y = \min(C,T)$ denote follow-up time; and $\Delta = I(T \leq C)$.  With this notation, note that when $\Delta=1$, $S_t=0$ for all $t<Y$, $S_t=1$ for all $t \geq Y$, and $C_t=0$ for all $t \leq Y$; when $\Delta=0$, $C_t=0$ for all $t<Y$, $C_t=1$ for all $t \geq Y$, and $S_t=0$ for all $t \leq Y$.  Thus, knowledge of $(Y,\Delta)$ provides information about the $S_t$ and $C_t$ processes.  

Let $Z^{(1)}_t$ denote the venlafaxine formulation on hand on day $t$ (0 for none, 1 for IM brand, 2 for IM generic, 3 for other). In the claims data we do not have access to what the patient actually has on hand, but we assume that patients take their drugs as instructed. Therefore, the value of $Z^{(1)}_t$ can be inferred based on the days supply of the last prescription fill of venlafaxine and the number of days since the fill took place. For example, if the patient fills a prescription for a $d$ days supply of ERC brand on day $t$, then $Z^{(1)}_t,...,Z^{(1)}_{t+d-1} = (3,...,3)$ for that patient. However, if the patient fills a new prescription for a different form of venlaxine before day $t+d-1$, then we assume that they have quit taking the old medication (that was filled on day $t$) and started taking the new one. On the other hand, if they fill again for the \textit{same} form of venlafaxine before day $t+d-1$, then we assume that the days supplies of both fills add up. That is, if the days supply of the second fill is $d'$, then for such a patient $Z^{(1)}_t,...,Z^{(1)}_{t+d+d'-1} = (3,...,3)$. Note that in our setting, $Z^{(1)}_1$ is always equal to either 1 or 2, depending on whether the patient initiated before or after generic became available. 

Let $Z^{(2)}_t$ denote the cumulative out-of-pocket costs associated with venlafaxine fills that have been made by $t$ (including formulations that are not IM brand or generic), and $Z_t = (Z^{(1)}_t,Z^{(2)}_t)$ denote the bi-variate time-varying exposure status. As discussed above, $Z^{(2)}_t$ is not the same variable as $\texttt{oop}_{\texttt{t}}$. We will be adjusting for $\texttt{oop}_{\texttt{t}}$ as a time-varying confounder, while treating $Z^{(2)}_t$ as an exposure because it is used to define adherence to the regimes in the ideal trial, and therefore the counterfactual outcomes of interest. 

Let $\bar{Z}_t$ denote the observed exposure history $(Z_1,...,Z_t)$, and $\bar{z}_t = (z_1,...,z_t)$ denote a fixed exposure history.  Let $S_t(\bar{z}_t)$ denote the counterfactual indicator of whether failure has occurred by day $t$ under the exposure history $\bar{z}_t$. For convenience, we use $\bar{z}$ to denote an entire fixed history of exposures. That is, $\bar{z} = \bar{z}_{t^*}$, where $t^*$ is the smallest $t$ such that $S_t(\bar{z}_t) = 1$. Let $T(\bar{z})$ denote the counterfactual failure time under exposure history $\bar{z}$. 

Let $g(t) = \sum_{k=1}^{t} p I\{\frac{k-1}{30} \text{ is an integer} \}$ be a step function denoting the total cost accumulated for filling venlafaxine by day $t$ under the ideal interventions, where $p$ is the fixed out-of-pocket cost of each prescription filled every 30 days as discussed in the ideal trial. A patient is considered adherent to the brand and generic regimes if $Z_t = (1,g(t))$ and $Z_t = (2,g(t))$, respectively, for all $t$ until failure occurs or the end of follow-up at 9 months. We will denote adherence to brand and generic on day $t$ with $Z_t = B_t$ and $Z_t = G_t$, respectively.

Let $U$ denote initiation date, and $u^*$ be the date of the first initiation to generic. In our database $u^*$ is August 8th, 2006. Define the \textit{brand era} to be the period of initiation times before $u^*$ and the \textit{generic era} to be the period of initiation times after and including $u^*$. Let $\gamma^{\bar{z}}_u(t) := P(T(\bar{z}) > t | U=u)$ be the probability of survival past day $t$ under intervention $\bar{z}$ conditional on initiation at $U=u$. 

Let $L_t$ be a vector of time-varying covariates measured prior to $S_t, C_t$ and $Z_t$ on day $t$ (i.e., $\texttt{rxb}_{\texttt{t}}$ and $\texttt{oop}_{\texttt{t}}$), where $L_1$ also contains baseline covariates (i.e., \texttt{race}, \texttt{sex}, \texttt{age}, \texttt{cci}, \texttt{ses}, \texttt{out}). We use $\bar{L}_t$ to denote the entire history of covariates up to day $t$.

\section{Challenges}
There are three major issues that must be addressed to identify $\gamma^{\bar{B}}_u(t)$ and $\gamma^{\bar{G}}_u(t)$ from the observed data: non-adherence to the interventions defined in the ideal clinical trial above, secular trends in health outcomes, and lack of treatment overlap. We discuss the issue of secular trends and non-overlap first. 

\subsection*{Secular Trends and Non-overlap}
Patients do not have access to generic until it becomes available, which means that treatment status is impacted by the date of initiation. If initiation time also impacts the outcome of interest (via a secular trend), this will result in temporal confounding. It is therefore necessary to adjust for initiation time in any study comparing brand to generic drugs where secular trends in the outcome are likely to be present. 

\begin{figure}[!h]
  \includegraphics[width=\linewidth]{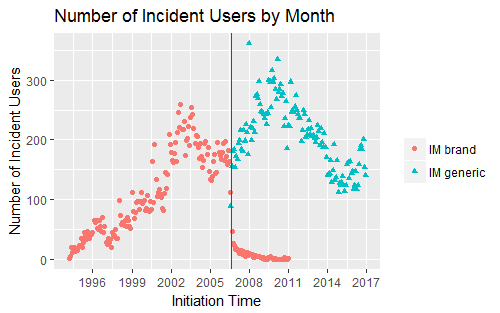}
  \caption{Incident use by month of brand and generic versions of immediate release venlafaxine. We have drawn a vertical red line at the date of first generic initiation (August 8th, 2006).}
  \label{fig:ven_use}
\end{figure}

Adjusting for initiation time with techniques that use standardization or propensity scores requires all levels of the treatment to have positive probability within every observed strata of initiation time. This requirement is known as \textit{positivity}, and it is violated in any study comparing brand to generic drugs, because generic use is impossible until the generic version is available. One possibility is to only compare brand to generic among patients initiating when both brand and generic are available; however, in our setting this is not feasible because there is almost no brand initiation once the generic becomes available (see Figure \ref{fig:ven_use}).  

\begin{figure}[!h]
  \includegraphics[width=\linewidth]{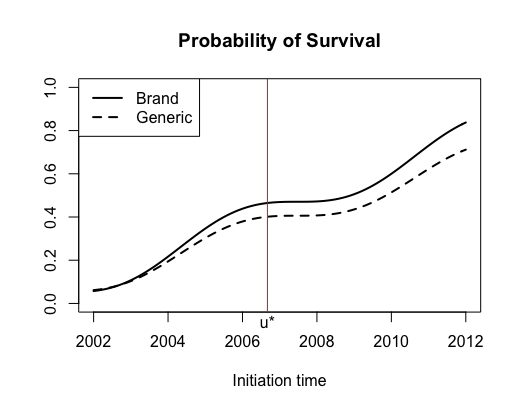}
  \caption{Hypothetical values of probability of survival past day $t$ conditional on treatment initiation time. The date of first generic initiation is represented by $u^*$.}
  \label{fig:TimeConfounding}
\end{figure}

Figure \ref{fig:TimeConfounding} illustrates temporal confounding by plotting hypothetical population values of $\gamma^{\bar{B}}_u(t)$ (solid line) and $\gamma^{\bar{G}}_u(t)$ as a function of $u$ at a fixed value of $t$. Assume, for now, that $\gamma^{\bar{B}}_u(t)$ and $\gamma^{\bar{G}}_u(t)$ are identifiable (i.e., estimable from the observed data) in their respective eras. 

Note that the curves are not flat, indicating the presence of a secular trend. Importantly, $\gamma^{\bar{B}}_u(t) > \gamma^{\bar{G}}_u(t)$ for all $u$, indicating that generic performs worse than brand at all values of $u$. However, assuming that $U$ is distributed uniformly over the time span of the plot, the overall probability of survival under brand use in the brand era is about $50\%$ lower than the probability of survival under generic use in the generic era. Therefore, a naive approach in which we estimate $\gamma^{\bar{B}}_{u}(t)$ in the brand era and compare it to an estimate of $\gamma^{\bar{G}}_u(t)$ in the generic era would lead to the wrong conclusion. This bias is due to temporal confounding. 

Importantly, note that an analysis which adjusts for $U$ to obtain a causal effect marginal over $U$ is impossible, because there is no overlapping period of initiation times when both $\gamma^{\bar{G}}_u(t)$ and $\gamma^{\bar{B}}_u(t)$ are identified. Our solution will be to extend regression discontinuity to whole survival curves in order to identify the parameters conditional on $U=u^*$, that is, $\gamma^{\bar{G}}_{u^*}(t)$ and $\gamma^{\bar{B}}_{u^*}(t)$. A comparison of these curves gives an effect that has a meaningful causal interpretation. However, the result is tied to the population of initiators at $U=u^*$, and so cannot be generalized to patients initiating at different times.

\begin{figure}[!h]
  \includegraphics[width=\linewidth]{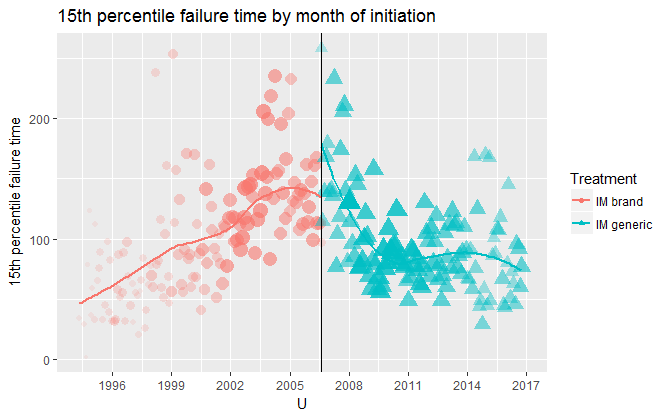}
  \caption{15th percentile survival times based on unadjusted Kaplan-Meier curves fit within each month of initiation time $U$.}
  \label{fig:secular}
\end{figure}

To determine whether there is a secular trend that might cause temporal confounding in our claims data we constructed Figure \ref{fig:secular}, which presents $15$th percentile survival times based on Kaplan-Meier curves estimated for each month of initiation time $U$. The size and transparency of each point represents the number of initiators for that month. Note the secular trend captured by the loess line. This suggests that unless we apply the method of regression discontinuity we may get a biased result.

\subsection*{Lack of Adherence}
So far we assumed that $\gamma^{\bar{B}}_u(t)$ and $\gamma^{\bar{G}}_u(t)$ are identified in the brand and generic eras, respectively. Here, we discuss how to identify them from the observed data, which features non-adherence to the interventions in the ideal trial discussed above.

Remember that patients are considered adherent when $Z_t = (1,g(t))$ (for brand) or $Z_t = (2,g(t))$ (for generic), until failure occurs or the end of follow-up at 9 months. There are three ways in which patients in the observed data can fail to adhere to the ideal interventions: (1) they fail to have the correct value of $Z^{(1)}_t$, for example by switching between different formulations of venlafaxine (e.g., IM vs. ERC), taking both brand and generic versions throughout their follow-up, or not having any prescription for venlafaxine on hand during parts of their follow-up; (2) they fail to have $Z^{(2)}_t = g(t)$, for example, because filling a prescription for brand has higher out of pocket costs than for generic, or prescription fills are not made at consistent intervals (e.g., every 30 days); and (3) $C < T$, that is, patients are lost to follow-up due to changes in enrollment before failure occurs. 

This leads to three potential sources of bias: (1) time-varying covariates that impact, and are impacted by, the daily exposure (the venlafaxine formulation on hand, whether it is brand or generic, and the accumulated out-of-pocket costs) can lead to time-varying confounding;  (2) a difference in the out of pocket cost associated with filling each prescription could result in a difference in survival time, meaning the effect size could be contaminated by an impact of cost (e.g., patients accumulating more costs may be more likely to switch to another cheaper anti-depressant, thereby decreasing failure time compared to patients accumulating fewer costs); and (3) informative censoring can lead to selection bias. 

In order to estimate the counterfactual survival curves under full adherence, we apply a method known as G-computation \cite{robins2008estimation}. Before discussing this further we first discuss time-varying confounding in more detail.

In our context, time-varying confounding can occur if $L_t$ (measured prior to $Z_t$  among those with $S_{t-1}=0$) has the following three properties:
\begin{enumerate}
	\item{$L_t$ is associated with $T$ (representable by $S_t,\ldots,S_{270}$), either because of some unmeasured shared cause $W$ or because of a direct impact}
	\item{$L_t$ causally impacts the exposures $Z_t$ }
	\item{$L_t$ is causally impacted by previous exposures $\bar{Z}_{t-1}$}
\end{enumerate}

To illustrate, consider the time varying covariate $\texttt{oop}_{\texttt{t}}$. Total out of pocket pharmacy costs may be associated with the outcome, since patients with higher pharmacy costs may be more likely to switch to a cheaper alternative (i.e., non-venlafaxine) drug than patients with low costs (fulfilling condition 1). Patients with high pharmacy costs may also be more likely to fill a prescription for generic venlafaxine, or none at all (fulfilling condition 2). Finally, having higher values of $Z^{(2)}_{t-1}$ may cause patients to not fill their other prescriptions, reducing their total out of pocket pharmacy costs (fulfilling condition 3). Failing to adjust for $L_t$ in this case could result in comparing two groups that differ in total out of pocket pharmacy costs, thereby biasing the results. 

\begin{figure}
\centering
\begin{tikzpicture}[>=stealth, node distance=2cm]
    \tikzstyle{format} = [draw, very thick, circle, minimum size=1.0cm,	inner sep=2pt]
    \tikzstyle{unobs} = [draw, very thick, red, circle, minimum size=1.0cm, inner sep=2pt]
	\begin{scope}
		\path[->, very thick]
			node[format](Z1){$Z_1$}
			node[format, right of=Z1](S1){$S_1$}
			node[format, right of=S1](Z2){$Z_2$}
			node[format, right of=Z2](S2){$S_2$}
			
			node[format, below of=Z1](L1){$L_1$}
			node[format, below of=Z2](L2){$L_2$}
			node[format, below of=L1, dashed](W){$W$}
			
			(Z1) edge (S1)
			(Z1) edge[bend left] (Z2)
			(Z1) edge[bend left] (S2)
			(Z1) edge (L2)
			
			(S1) edge[bend left] (S2)
			(S1) edge (Z2)
			(S1) edge (L2)
			
			(L1) edge (Z1)
			(L1) edge (Z2)
			(L1) edge (L2)
			
			(Z2) edge (S2)
			
			(L2) edge (Z2)
			(W) edge (L1)
			(W) edge (S1)
			(W) edge (L2)
			(W) edge[bend right=45] (S2)

			;
	
	\end{scope}
\end{tikzpicture}
\hspace{1cm}
\raisebox{3em}{
\begin{tikzpicture}[>=stealth, node distance=2cm]
    \tikzstyle{format} = [draw, very thick, circle, minimum size=1.0cm,	inner sep=2pt]
    \tikzstyle{unobs} = [draw, very thick, red, circle, minimum size=1.0cm, inner sep=2pt]
	\begin{scope}
		\path[->, very thick]
			node[format](Z1){$Z_1$}
			node[format, right of=Z1](S1){$S_1$}
			node[format, right of=S1](Z2){$Z_2$}
			node[format, right of=Z2](S2){$S_2$}
			
			node[format, below of=Z1](L1){$L_1$}
			node[format, below of=Z2](L2){$L_2$}
			
			(Z1) edge (S1)
			(Z1) edge[bend left] (Z2)
			(Z1) edge[bend left] (S2)
			(Z1) edge (L2)
			
			(S1) edge[bend left] (S2)
			(S1) edge (Z2)
			(S1) edge (L2)
			
			(L1) edge (Z1)
			(L1) edge (S1)
			(L1) edge (Z2)
			(L1) edge (L2)
			(L1) edge (S2)
			
			(Z2) edge (S2)
			
			(L2) edge (Z2)
			(L2) edge (S2)

			;
	
	\end{scope}
\end{tikzpicture}
}
\caption{Directed acyclic graphs (DAGs) depicting time-varying confounding when the association between $L_2$ and $S_2$ is due to an unmeasured shared cause, $W$ (on the left) and when the association between $L_2$ and $S_2$ is due to a direct causal path from $L_2$ to $S_2$ (on the right).}
\label{fig:tvc1}
\end{figure}
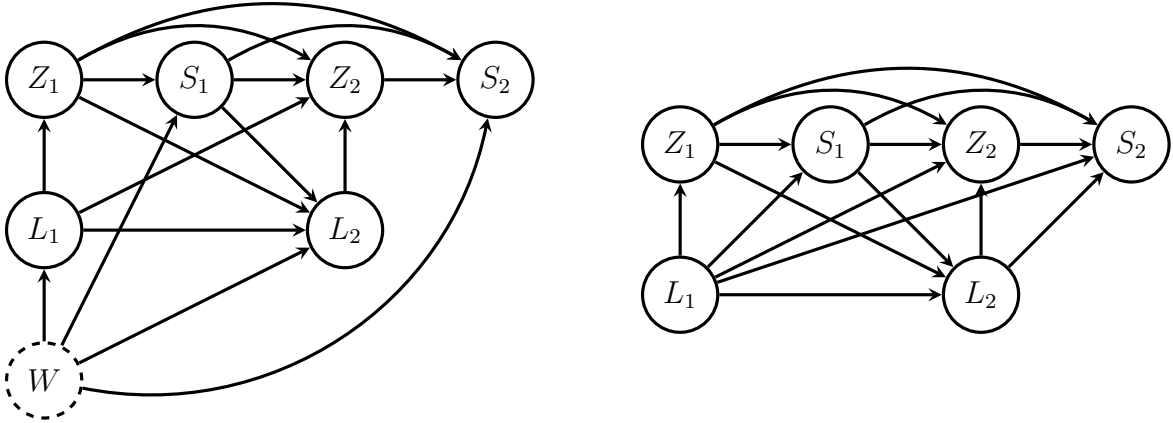

The problem is that standard methods to adjust for confounding may not work well with time-varying confounding. To see this, examine the directed acyclic graphs (DAGs) in Figure \ref{fig:tvc1}, which depict the two different ways the first condition on $L_t$ can be met. For expository purposes, we only show graphs for $t=1,2$. Suppose an analyst uses a discrete-time logistic hazards regression model to evaluate the effect of $Z_1$ and $Z_2$ on $T$ using $L_t$ as a time-varying covariate.  The analyst would be modeling the conditional hazard at time 2: $P(S_2=1|S_1=0,Z_1,Z_2,L_1,L_2)$. If the true causal structure is depicted by the DAG in Figure \ref{fig:tvc1} on the left, meaning the association between $L_2$ and $S_2$, conditional on $S_1=0$, is due to the unmeasured shared cause $W$, then $L_2$ is a collider on the path $Z_1 \rightarrow L_2 \leftarrow W \rightarrow S_2$. In that case, conditioning on the collider $L_2$ can bias the estimated effect of $Z_1$ on $S_2$ in the hazard model \cite{cole2009illustrating}. On the other hand, if the true causal structure is depicted by the DAG on the right, meaning the association, conditional on $S_1=0$, between $L_2$ and $S_2$ is due to a direct causal path from $L_2$ to $S_2$, then $L_2$ is a mediator of the effect of $Z_1$ on $S_2$. In that case, conditioning on the mediator $L_2$ can block some of the effect of $Z_1$ on $S_2$ in the hazard model. In either case, we may not be able to properly understand the joint effect of $Z_1$ and $Z_2$ on $T$ using hazard modeling. (See \citeA{daniel2013methods} for discussion of time varying confounders.)

G-computation allows us to iteratively adjust for $L_t$ without inducing bias by blocking mediation paths or conditioning on a collider. It also allows us to adjust for the impact of  censoring. (See \citeA{robins2008estimation} for a discussion of using G-computation with time-varying exposures. See \citeA{wang2011estimating} for an application of G-computation in a context that is similar to ours.)

\section{Methods}

\subsection{Causal Model}
The target parameters are the conditional counterfactual survival curves $\gamma^{\bar{B}}_{u^*}(t)$ and $\gamma^{\bar{G}}_{u^*}(t)$. Here we formally state the assumptions we make in order to identify them from the observed data.
\subsection*{Assumptions}
\begin{assumption} (Sequential Randomization)  For $\bar{z} = \bar{B},\bar{G}$ and $t=1,...,\min\{270,T(\bar{z})\}$,
\[
Z_t \perp T(\bar{z}) | U, \bar{L}_t, \bar{Z}_{t-1} = \bar{z}_{t-1}, S_{t-1} = C_{t-1} =0.
\]
\end{assumption}
\noindent This assumption states that, conditional on initiation time, covariate history, adherence to $\bar{z}$ through day $t-1$ and surviving uncensored to day $t$, exposure at day $t$ is independent of the potential outcome $T(\bar{z})$. 

\begin{assumption} (Non-informative Censoring) For $\bar{z} = \bar{B},\bar{G}$ and $t=1,...,\min\{270,T(\bar{z})\}$,
\[
C_{t} \perp T(\bar{z}) | U, \bar{L}_t, \bar{Z}_{t} = \bar{z}_t, S_{t-1} = C_{t-1} = 0.
\]
\end{assumption}
\noindent This assumption states that, conditional on initiation time, covariate history, adherence to $\bar{z}$  through day $t-1$ and surviving uncensored to day $t$, censoring at day $t$ is independent of the potential outcome $T(\bar{z})$.

\begin{assumption} (Consistency)  For $\bar{z} = \bar{B},\bar{G}$ and $t=1,...,\min\{270,T(\bar{z})\}$,
\[
S_t(\bar{z}_{t}) = S_t \text{ if } \bar{Z}_t = \bar{z}_{t} \text{ and } C_t = S_{t-1}=0.
\]
\end{assumption}
\noindent This assumptions states that, within the first 9 months of initiation, if we actually observe no censoring and adherence to $\bar{z}$ through day $t$ then we observe the potential outcome $S_t(\bar{z}_{t})$.

For the next assumption, let $h_t(\cdot)$ be the joint density of $(U,\bar{L}_t,\bar{Z}_{t-1},C_{t-1}=S_{t-1}=0)$, and $h_{Z_t}(\cdot| U, \bar{L}_t, \bar{Z}_{t-1})$ be the conditional density of $Z_t$ given $(U, \bar{L}_t, \bar{Z}_{t-1}, C_{t-1} = S_{t-1}=0)$. We drop $\bar{Z}_{t-1}$, $C_{t-1}$, and $S_{t-1}$ from these definitions when $t=1$.

\begin{assumption} (Era-Specific Positivity) If $u < u^*$, then for $t=1,...,\min(270,T(\bar{B}))$, 
\[
h_{Z_t}(B_t | u, \bar{l}_t, \bar{B}_{t-1}) > 0 \text{ if }h_t(u,\bar{l}_t,\bar{B}_{t-1}) > 0.
\] 
If $u \ge u^*$, then for $t=1,...,\min(270,T(\bar{G}))$, 
\[
h_{Z_t}(G_t | u, \bar{l}_t, \bar{G}_{t-1}) > 0 \text{ if }h_t(u,\bar{l}_t,\bar{G}_{t-1}) > 0.
\]
\end{assumption}
\noindent This assumption states that any patient who has initiated in the brand (or generic) era and remained uncensored and adherent through day $t-1$ has a positive probability of continued adherence to brand (or generic) on day $t$ within the first 9 months of initiation. 

For the next assumption, let $P(C_t = 0 | U, \bar{L}_t, \bar{Z}_t)$ denote the conditional probability of $C_t = 0$ given $(U, \bar{L}_t, \bar{Z}_t, C_{t-1} = S_{t-1} = 0)$. Again, we drop $\bar{Z}_t, C_{t-1}$, and $S_{t-1}$ when $t=1$.

\begin{assumption} (C-Positivity) For $\bar{z}=\bar{B},\bar{G}$ and $t=1,...,\min(270,T(\bar{z}))$, 
\[
P(C_t = 0 | u, \bar{l}_t, \bar{z}_t) > 0 \text{ if } h_t(u,\bar{l}_t,\bar{z}_{t-1}) \cdot h_{Z_t}(z_t | u, \bar{l}_t, \bar{z}_{t-1}) > 0.
\]
\end{assumption}
\noindent This assumption states that any patient who has remained uncensored through day $t-1$ and adherent through day $t$, has a positive probability of remaining uncensored through day $t$ within the first 9 months of initiation.

The previous 5 assumptions are required to perform G-computation. The next is required to extend the framework of regression discontinuity to survival curves.

\begin{assumption}(Continuity) For $t \le 270$, 
\[
\lim_{u \to u^{*-}} \gamma^{\bar{B}}_u(t) = \gamma^{\bar{B}}_{u^*}(t)
\]
and 
\[
\lim_{u \to u^{*+}} \gamma^{\bar{G}}_u(t) = \gamma^{\bar{G}}_{u^*}(t).
\]

\end{assumption} 
\noindent This assumption states that the counterfactual probability of failure before any day within the first 9 months under continuous treatment with either brand or generic is left- or right-continuous at the date of generic market entry, $u^*$, respectively. This amounts to assuming no sudden shifts in the counterfactual probability of survival as a function of initiation time near $u^*$, caused by, for example, a sudden shift in the population characteristics of those initiating to IM venlafaxine around that time.

\subsection*{Identification}

Under Assumptions $1$ to $5$, we can prove the following lemma by applying G-computation (see Robins (1986) for proof). We use $F_k(.|U,\bar{L}_{t-1}, \bar{Z}_{t-1})$ to denote the conditional distribution of $L_k$ given $(U,\bar{L}_{k-1}, \bar{Z}_{k-1}, C_{k-1} = S_{k-1} = 0)$, and $P(S_k = 0 | U, \bar{L}_k, \bar{Z}_k)$ to denote the conditional probability of $S_k=0$ given $(U, \bar{L}_k, \bar{Z}_k, C_k = S_{k-1} = 0)$. Once again, we drop $\bar{L}_{k-1}, \bar{Z}_{k-1}, C_{k-1}$, and $S_{k-1}$ from these definitions when $k=1$. Note also we do not include $C_{k-1}$ and $S_{k-1}$ in the notation for the set of variables we condition on, since in all cases we are conditioning on $C_{k-1} = S_{k-1} = 0$.
\begin{lemma} 
For $u < u^*$ and all $t \le 270$,
\begin{align*} 
	\gamma_u^{\bar{B}}(t) = \int_{l_1} &... \int_{l_t} \prod_{k=2}^{t} \{P(S_k = 0| u, \bar{l}_k, \bar{B}_k)dF_k(l_k|u,\bar{l}_{k-1}, \bar{B}_{k-1})\}P(S_1=0|u, l_1, B_1)dF_1(l_1|u).
\end{align*}

And for $u \ge u^*$ and all $t \le 270$,
\begin{align*} 
	\gamma_u^{\bar{G}}(t) = \int_{l_1} &... \int_{l_t} \prod_{k=2}^{t} \{P(S_k = 0| u, \bar{l}_k, \bar{G}_k)dF_k(l_k|u,\bar{l}_{k-1}, \bar{G}_{k-1})\}P(S_1=0|u, l_1, G_1)dF_1(l_1|u).
\end{align*}
\end{lemma}

Lemma 5.1 implies both $\gamma^{\bar{B}}_u(t)$ and $\gamma^{\bar{G}}_u(t)$ are identifiable in the brand and generic eras, respectively. Applying Assumption $6$ (continuity) allows us to take limits in order to identify $\gamma_{u^*}^{\bar{B}}(t)$ and $\gamma_{u^*}^{\bar{G}}(t)$. 

\subsection{Estimation and Inference}
We assume that we observe $n$ i.i.d copies of $(U, Y, \Delta, \bar{L}_Y,\bar{Z}_Y)$. To estimate $\gamma_{u^*}^{\bar{B}}(t)$ and $\gamma_{u^*}^{\bar{G}}(t)$ we propose models for $P(S_k = 0 | U, \bar{L}_k, \bar{Z}_k)$ (i.e., the conditional distribution of $S_k$ given $U,\bar{L}_k, \bar{Z}_k, C_k = S_{k-1} = 0$) and $F_k(.|U,\bar{L}_{t-1}, \bar{Z}_{t-1})$ (i.e., the conditional distribution of $L_k$ given $U,\bar{L}_{k-1}, \bar{Z}_{k-1}, C_{k-1} = S_{k-1} = 0$) that are smooth in $U$ near $u^*$, for k=1,...,t. We then fit those models and numerically approximate the integrals in Lemma 5.1 using Algorithm 1 below. We do this for $t = 1,...,270$ to estimate 9 month survival curves. We then estimate 95\% confidence intervals of the difference restricted mean survival using non-parametric bootstrap. 

\begin{algorithm}[!h]
\caption{Numerical Integration.} 
\label{alg:alg}
\begin{algorithmic}
\STATE $m\gets 1$
\WHILE {$m \le M$}
	\STATE{Sample $L_1^{(m)} \sim \hat{F}_1(\cdot | u^*)$}
	\STATE{Sample $S_1^{(m)} \sim \hat{P}(\cdot | u^*, L_1^{(m)}, Z_1 = z_1)$}
	\STATE {$k \gets 2$}
	\WHILE {$k \le t$}
		\STATE {Sample $L_k^{(m)} \sim \hat{F}_k(\cdot | u^*, \bar{L}^{(m)}_{k-1}, \bar{Z}_{k-1} = \bar{z}_{k-1})$}
		\IF{$S_{k-1}^{(m)} = 0$}
			\STATE {Sample $S_{k}^{(m)} \sim \hat{P}(\cdot | u^*, \bar{L}_k^{(m)}, \bar{Z}_k = \bar{z}_k)$}
		\ELSE
			\STATE {$S_{k}^{(m)} \gets 1$}
		\ENDIF
	\ENDWHILE
	\STATE {$m \gets m+1$}
\ENDWHILE
\STATE {$\hat{\gamma}^{\bar{z}}_{u^*}(t) \gets 1 - \frac{1}{M} \sum_{m-1}^M S_t^{(m)}$}
\end{algorithmic}
\end{algorithm}

We were restricted to conducting all analyses on a server provided by OptumLabs which had limited space and only 4 processing cores. Therefore we chose to split the data into 3 random partitions and perform the entire analysis on each partition in parallel. On each partition, we estimate $\gamma_{u^*}^{\bar{B}}(t)$ and $\gamma_{u^*}^{\bar{G}}(t)$, along with bootstrapped point-wise standard errors of $\log(\hat{\gamma}_{u^*}^{\bar{B}}(t))$ and $\log(\hat{\gamma}_{u^*}^{\bar{G}}(t))$, using $100$ re-samples (of size equal to the sample on that partition) for the bootstrap. We average these 3 estimates and compute the overall standard error as $SE = \frac{1}{3}\sqrt(\sum_{p=1}^3 SE_p^2)$, where $SE_p$ is the bootstrapped standard error on partition $p$. 

We set the simulation size for numerical integration to $M = 40$,$000$ on each partition, sampled from $\hat{F}_1(\cdot | u^*)$. To sample from $\hat{F}_1(\cdot | u^*)$ we take a weighted sample with replacement from the observed baseline values, with sampling weights given by $\phi(\frac{U - u^*}{h})$. The choice of bandwidth $h$ determines the dates between which 95\% of the units will be sampled. For example, setting $h=365$ days ensures that 95\% of the patients sampled initiate treatment within two years of $u^*$. Note that we cannot simply take an unweighted sample with replacement of baseline values of $L_1$, because this would be a sample from $\hat{F}_1(\cdot)$ not $\hat{F}_1(\cdot | u^*)$.

\section{Simulation Study}
To characterize the degree of bias induced by temporal and time-varying confounding, we compared the performance of our proposed method with three others: one that adjusts for temporal but not time-varying confounding, one that adjusts for time-varying but not temporal confounding, and one that adjusts for neither. Moreover, we compared these four methods in two different scenarios: one in which the relationship between $U$ and the probability of survival is modeled correctly, and one in which it is modeled incorrectly but flexibly with natural cubic spline terms.

To match the complexity of the data we simulated data sets from models fit to the actual data. We modeled:
\begin{itemize}
\item[(1)] $S_t | U, L^*_t, Z_t$
\item[(2)] $Z_t | U, L^*_t$
\item[(3)] $L^*_t | L^*_{t-1}, Z_{t-1}$,
\end{itemize}
where we let $L^*_t$ simply be an indicator of whether the prescription burden on day $t$ was zero or greater than zero. The dependence of $S_t$ and $Z_t$ on $U$ ensures the presence of temporal confounding, and the dependence of $L^*_t$ on $Z_{t-1}$ and $Z_t$ on $L^*_t$ ensures the presence of time-varying confounding. 

One challenge with modeling exposure ($Z_t$) in a realistic manner is that there are stretches of time during which part of it, $Z_t^{(1)}$, is deterministic given the past. If a patient fills a prescription for a 30 days supply of IM brand on day $t$, then their value of $Z_t^{(1)}$ will be set at 1 from days $t$ to $t+29$. To mimic this structure in the simulated data sets, we only modeled exposure at \textit{fill opportunities}. Day 1 of follow-up is the first fill opportunity, and by definition each patient uses that opportunity to fill a prescription for either IM brand or IM generic. The next fill opportunity is determined by the days supply of the first fill, where we assume that patients do not have an opportunity to fill again until the days supply has been exhausted. At the next fill opportunity, the patient can fill a prescription for IM brand or generic again, for some other version of venlafaxine, or choose not to fill a prescription for any form of venlafaxine at all. In the latter two cases, the patient is non-adherent. In the last case (no fill), the patient does not accumulate any out of pocket costs associated with treatment with venlafaxine. 

Let $D_t$ and $Z_t^{(2)*}$ be the days supply and out of pocket cost of the fill of venlafaxine made on day $t$, assuming $t$ is a fill opportunity day. We only model $Z_t | U, L^*_t$ if $t$ is a fill opportunity, and split it into models for 
\begin{itemize}
\item[(2.1)] $Z_t^{(1)} | U, L^*_t$
\item[(2.2)] $D_t | Z_t^{(1)}$
\item[(2.3)] $Z_t^{(2)*} | Z_t^{(1)}, D_t$
\end{itemize}
When simulating from these models, if the value of $Z_t^{(1)}$ is 0, then $D_t$ is forced to 1 (meaning the next day is another opportunity for the patient to fill) and $Z_t^{(2)*}$ is forced to 0. If $Z_t^{(1)}$ is any other value than 0 and $D_t > 1$, then we force $(Z_{t+1}^{(1)},...,Z_{t+D_t -1}^{(1)}) = (Z_{t}^{(1)},...,Z_{t}^{(1)})$ and $(Z_{t+1}^{(2)*},...,Z_{t+D_t -1}^{(2)*}) = (0,...,0)$. $Z_t^{(2)}$ is then calculated as $\sum_{k=1}^t Z_k^{(2)*}$.

In both scenarios (the secular trend known and unknown), we set the true models for (1) through (3) by fitting models to the real data: logistic regression for $S_t$ and $L^*_t$, multinomial regression for $D_t$ and $Z_t^{(1)}$, and linear regression (using log transformed values) for $Z_t^{(2)*}$. We fit separate models for brand era and generic era initiators. For the scenario when the secular trend is unknown, we defined the true relationship between $U$ and $S_t$ using a sine wave. That is, $\text{logit} P(S_t = 1 | U, L^*_t, Z_t) = \beta_0 + \beta_1L^*_t + \beta_2 Z_t + \beta_3 \sin(U/\bar{U})$, where $\bar{U}$ is the mean of $U$. This led to a true restricted mean difference of 6.77 days in the scenario with known (and linear) secular trend, and -51.05 days in the scenario with unknown secular trend. 

We simulated 1000 data sets with a sample size of 1000 under each scenario and analyzed the data. The results of the simulation study are given in Table \ref{table:sim}. Note that the analysis with the least bias in both scenarios is the one that adjusts for both sources of confounding. Importantly, when the form of the secular trend is not known, the use of flexible natural cubic spline terms appears to keep the bias low, while failing to address the temporal confounding at all leads to very large bias. Note also that the root mean squared errors are much higher in this context. For the analyses that did not address temporal confounding, this is mostly due to the large bias; but for the other analyses this is due to the increased variance associated with estimating the secular trend. 
\begin{table}
\centering
\begin{tabular}{ccccccc}
\hline\hline
& \multicolumn{3}{c}{Secular Trend Modeled Correctly} &\multicolumn{3}{c}{Secular Trend Modeled Incorrectly} \\
Analysis & Bias & $\sqrt{MSE}$ & 95\% CI Coverage & Bias & $\sqrt{MSE}$ & 95\% CI Coverage\\
\hline
Both & -0.48 & 6.50 & 94.40\% & -7.57 & 18.90 & 91.80\% \\
Time-varying & -3.33 & 6.80 & 90.30\% & -58.40 & 58.60 & 0\% \\
Temporal & -1.58 & 6.40 & 94.40\% & -11.060 & 21.20 & 89.60\% \\
Neither & -4.67 & 7.40 & 86.3\% & -59.87 & 60.10 & 0\% \\
\hline
\end{tabular}
\caption{Simulation study results in two scenarios, comparing results of the analysis that adjusts for both temporal and time-varying confounding, only time-varying confounding, only temporal confounding, and neither sources of confounding.}
\label{table:sim}
\end{table}

\section{Analysis}
In what follows, let $X$ be the vector of baseline covariates (\texttt{age}, \texttt{race}, \texttt{sex}, \texttt{ses}, \texttt{cci}, \texttt{out}), so that $L_1 = (\texttt{oop}_{\texttt{1}},\texttt{rxb}_{\texttt{1}},X)$. Let $\tilde{\texttt{oop}}_{\texttt{t}}$ denote the inverse hyperbolic sine transform of $\texttt{oop}_{\texttt{t}}$, which stabilizes the variance and preserves zero points. Define $\texttt{zero}_{\texttt{t}} = I(\texttt{oop}_{\texttt{t}} > 0)$. 

We fit the following Markovian models:
\begin{enumerate}
\item $\texttt{rxb}_{\texttt{t}}|U,{X},\texttt{rxb}_{\texttt{t-1}}, \tilde{\texttt{oop}}_{\texttt{t-1}}, {Z}_{{t}}, {S}_{{t-1}} = {C}_{{t}} = 0$
\item $\texttt{zero}_{\texttt{t}} |U,{X},\texttt{rxb}_{\texttt{t}}, \tilde{\texttt{oop}}_{\texttt{t-1}}, \texttt{zero}_{\texttt{t-1}}, {Z}_{{t}}, {S}_{{t}} = {C}_{{t}} = 0$ 
\item $\tilde{\texttt{oop}}_{\texttt{t}}^{-1} |U,{X},\texttt{rxb}_{\texttt{t}}, \tilde{\texttt{oop}}_{\texttt{t-1}},\texttt{zero}_{\texttt{t}} = 1, {Z}_{{t}}, {S}_{{t-1}} = {C}_t = 0$
\item ${S}_{{t}}|U,{X}, \texttt{rxb}_{\texttt{t}}, \tilde{\texttt{oop}}_{\texttt{t}}, \texttt{zero}_{\texttt{t}}, {Z}_{{t}}, {S}_{{t-1}} = {C}_{{t}} = 0$
\end{enumerate}
In all we treated \texttt{cci}, \texttt{out}, $\texttt{rxb}_{\texttt{t}}$, and \texttt{ses} as factors. We collapsed categories 0, 2, and 3 of the exposure $Z^{(1)}_t$ for brand initiators, and categories 0, 1, and 3 for generic initiators. We used natural cubic splines terms with 3 degrees of freedom for \texttt{age}, $\texttt{oop}_{\texttt{t}}$, and ${Z}^{(2)}_{{t}}$, and 5 degrees of freedom for $U$. We used ordinal logistic regression for model (1), logistic regression for (2) and (4), and gamma regression (with a log link) for (3). In all models we included natural cubic spline terms for $t$ with 3 degrees of freedom. To ensure robustness to the choice of $h$, we performed analyses at three values of $h$: 365, 730, and 2920 days, corresponding to 95\% of the baseline sample initiating within 2 years, 4 years and 16 years of $u^*$, respectively. 

We assessed goodness of fit of the models for the categorical variables $\texttt{rxb}_{\texttt{t}}$, $\texttt{zero}_{\texttt{t}}$, and $S_t$ by fitting them to one partition of the data and looking at their prediction error on another partition, where we took the estimated most probable value as the prediction. For the models fit to brand initiators, the incorrect category was predicted only $6.1\%$, $0.094\%$, and $0.12\%$ of the time, respectively. The error rates for those fitted to the generic initiators were $5.8\%$, $0.13\%$, and $0.13\%$, respectively.

For the gamma regression of $\tilde{\texttt{oop}}_{\texttt{t}}$, we fit the model to a random subset of patients including both brand and generic initiators (due to restrictions in computational resources) and generated 100 simulates for each observation using the model fit. We then calculated the empirical cumulative density function for each set of simulates, and determined its value at the observed value of $\tilde{\texttt{oop}}_{\texttt{t}}$. The cumulative density at an observed data value will be uniformly distributed if the model is correct. Kolmogorov's distance was computed to be 0.01. This procedure was implemented using the R package DHARMa \cite{dharma}.


\begin{figure}[!h]
  \includegraphics[width=\linewidth]{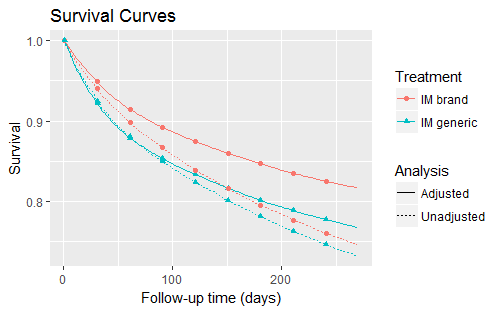}
  \caption{Estimated counterfactual survival curves $\hat{\gamma}_{u^*}^{\bar{B}}(t)$ and $\hat{\gamma}_{u^*}^{\bar{G}}(t)$, as well as unadjusted Kaplan-Meier curves.}
  \label{fig:curves}
\end{figure}

Figure \ref{fig:curves} shows the estimated curves $\hat{\gamma}_{u^*}^B(t)$ and $\hat{\gamma}_{u^*}^G(t)$ with choice of $h$ set at 365 days, as well as unadjusted Kaplan-Meier curves for comparison. Note that the adjusted curves are higher than the unadjusted curves, suggesting that adherence to $\bar{B}$ and $\bar{G}$ decreases the probability of failure. The estimated restricted mean difference is $10.54$ days with a standard error of $9.58$, resulting in a 95\% CI of (-8.62, 29.70).  The estimated restricted mean difference is $11.53$ days (with a standard error of 8.88) and $9.44$ days (with a standard error of 5.54), for bandwidth choices $730$ and $2920$, respectively. These results do not provide evidence of a difference in failure times between IM brand venlafaxine and its generic version.

\section{Discussion}

We developed a method that can be used by regulators to assess the therapeutic equivalence of brand and generic drugs using observational claims data. This required identification of the survival curves under the idealized interventions $\bar{B}$ and $\bar{G}$ in the presence of lack of adherence, secular trends, and treatment non-overlap. Under the assumptions discussed in Section 5, our approach overcomes these challenges. When we apply the method to IM venlafaxine, the results do not provide evidence for a lack of therapeutic equivalence of brand and generic.

We chose not to include patients initiating on brand in the generic era, which resulted in a sharp regression discontinuity design wherein the probability of being treated with brand jumps from 1 to 0 (see \citeA{imbens2008regression} for a discussion of the difference between sharp and fuzzy regression discontinuity). Given the relatively small fraction of patients initiating on brand in the generic era (around $0.9\%$), and the fact that the incidence of brand initiation quickly drops to zero, we considered the extra complication of incorporating these data into a fuzzy regression discontinuity design to be of marginal benefit. However, in some cases the incidence of brand initiation in the generic era will be high enough that excluding these patients could induce bias if they are systematically different from those initiating on generic in the generic era. More work is therefore needed to extend this method to a fuzzy regression discontinuity design. 

There are a number of limitations of our proposed method. First, the result is tied to initiation time $U=u^*$, making it difficult to generalize to patients initiating at different times. However, this is true of randomized clinical trials generally, as patients initiate treatment during a finite window of time. 

Second, our outcome does not encompass every aspect of therapeutic equivalence. Therapeutic equivalence entails that brand and generic do not differ in time to failure as it is defined in this study, however the converse does not hold. Therefore, this is not a complete assessment of therapeutic equivalence. 

Third, we did not have data on actual patient behavior. Therefore, we cannot estimate what would happen had patients continuously \textit{taken} brand or generic, rather than continuously filled their prescriptions. However, this is a limitation of the data rather than the method. 

Fourth, the method requires specifying fully parametric models for $P_k$ and $F_k$. Misspecification of these models could lead to biased results. Extensions of this method to handle flexible semiparametric or nonparametric models is therefore needed. At the very least, care should be taken to ensure the models are flexible and fit the data well.

Fifth, the causal interpretation relies on the untestable assumptions described in Section 5, such as sequential randomization. Whether the method is applicable in a given setting relies on the scientific plausibility of these assumptions.

Finally, we were restricted to conducting all analyses on a server provided by OptumLabs which had limited space and only 4 processing cores. This guided some of our analytic choices, such as the number of bootstrap samples, the choices of bandwidth ($h$), and the decision to partition the data.


\section*{Acknowledgments}

We would like to thank Ramin Mojtabai for providing his medical expertise, Vijay S. Nori for aiding in the computation, Michael A. Rosenblum and Ilya Shpitser for providing their insight on how to think about the problem from a causal inference perspective, and OptumLabs.

Funding for this manuscript was made possible, in part, by the Food and Drug Administration through grant 1U01FD005556-01. Views expressed in the manuscript do not necessarily reflect the official policies of the Department of Health and Human Services; nor does any mention of trade names, commercial practices, or organization imply endorsement by the United States Government. 

\bibliography{References}

\end{document}